\documentclass[%
amsmath,a4paper,amsfonts,amssymb,
aps,pra,onecolumn,longbibliography
]{revtex4-2}

\usepackage{graphicx}
\usepackage{dcolumn}
\usepackage{bm}
\usepackage{dsfont}
\usepackage{physics} 
\usepackage{upgreek} 
\usepackage[colorlinks=true,allcolors=blue]{hyperref}
\usepackage{xcolor}
\usepackage{url}
\usepackage{indentfirst}
\usepackage{multirow}
\usepackage{hhline}

\newcommand{\dL}{\mathrm{L}}
\newcommand{\doc}{\mathrm{oc}}
\newcommand{\dT}{T}
\newcommand{\dt}{\mathrm{tg}}
\newcommand{\dML}{\mathrm{ML}}
\newcommand{\dflip}{\mathrm{flip}}

\begin{document}

\title{Floquet operator engineering for quantum state stroboscopic stabilization}

\author{F.~Arrouas$^1$, N.~Ombredane$^1$, L.~Gabardos$^1$, E.~Dionis$^2$, N.~Dupont$^1$, J.~Billy$^1$, B.~Peaudecerf$^1$, D.~Sugny$^2$, D.~Gu\'ery-Odelin$^1$}

\email{dgo@irsamc.ups-tlse.fr}

\affiliation{
$^1$ Laboratoire Collisions Agr\'egats R\'eactivit\'e, UMR 5589, FERMI, UT3, Universit\'e de Toulouse, CNRS,\\
118 Route de Narbonne, 31062 Toulouse CEDEX 09, France \\
$^2$ Laboratoire Interdisciplinaire Carnot de Bourgogne, UMR 6303,\\
9 Avenue A. Savary, BP 47 870, F-21078 Dijon Cedex, France
}

\date{\today}

\begin{abstract}
Optimal control is a valuable tool for quantum simulation, allowing for the optimized preparation, manipulation, and measurement of quantum states. Through the optimization of a time-dependent control parameter, target states can be prepared to initialize or engineer specific quantum dynamics. In this work, we focus on the tailoring of a unitary evolution leading to the stroboscopic stabilization of quantum states of a Bose-Einstein condensate in an optical lattice.
We show how, for states with space and time symmetries, such an evolution can be derived from the initial state-preparation controls; while for a general target state we make use of quantum optimal control to directly generate a stabilizing Floquet operator.
 Numerical optimizations highlight the existence of a quantum speed limit for this stabilization process, and our experimental results demonstrate the efficient stabilization of a broad range of quantum states in the lattice.
\end{abstract}

\maketitle

\section{Introduction}

In any experimental platform that aims to perform quantum simulation, it is desirable to implement a wide range of control capabilities, allowing for the preparation and detection of diverse quantum states, and the implementation of various dynamics~\cite{Cirac2012}. Periodic modulation of parameters, or Floquet engineering, is an expanding direction of research that seeks to broaden the scope of quantum simulators, achieving effective Hamiltonians~\cite{Dalibard2011,goldman_dalibard_2014}, emulating dissipative environments~\cite{impens_2023}, or accessing synthetic dimensions~\cite{Ozawa2019}.

Bose-Einstein condensates (BEC) are very-well suited to quantum simulation, owing to the exquisite degree of control available in cold atoms experiments, and the possibility they offer to manipulate the collective macroscopic wavefunction of the condensate. Cold atoms can be placed in a wide range of potentials, an emblematic case being the periodic potentials generated with optical lattices; an extensive review on atomic gases in periodically driven optical lattices can be found in~\cite{Eckardt_2017}. 

In previous work~\cite{dupont_2021,dupont_2023}, we have demonstrated how the external quantum state of a BEC in a one-dimensional optical lattice could be optimally prepared in target states characterized by their momentum wavefunction, or their phase space distribution. This relied on an optimal modulation of a single parameter -- the lattice position, derived from quantum optimal control (QOC) theory~\cite{boscain_2021}. QOC has been applied successfully to quantum systems in various contexts, especially for quantum technology applications~\cite{koch_2022}, the most closely related to this work being quantum interferometry~\cite{saywell2020,frank2014,weidner2018,ledesma_2023} and quantum simulation~\cite{zhou2018,frank2016} with cold atoms.

While our previous work aimed at preparing specific, non-stationary quantum states starting from the ground state of the system, we extend in this work the scope of the optimal state-to-state manipulation to address the optimal transfer between two non-stationary states of the system. We focus on the optimized transfer of a non-stationary state to itself in a finite duration, \emph{i.e.} the stroboscopic stabilization of the state. This task effectively amounts to engineering an optimal Floquet operator for which the state of interest is an eigenstate. It can be seen as a promising new tool for the field of Floquet engineering, in which most studies only consider a periodic modulation with a simple single-frequency shape. Note that related optimal control problems have been recently formulated in the context of material science~\cite{Castro_22,Castro_2023} or to realize ``undo" operations on the internal state of cold atoms~\cite{mastroserio_2022}.

Using a squeezed Gaussian state as a case study, we show how such an optimal "stabilizing" Floquet operator may be derived from existing preparation controls by exploiting state symmetries, or can be obtained by a direct numerical optimization of the lattice phase modulation. The optimization of the Floquet operator for different control durations reveals the existence of a minimum time from which stroboscopic stabilization can be achieved. This minimum time can be interpreted as a quantum speed limit~\cite{Deffner2017,Caneva2009} to transform a quantum state into itself. Physically, it corresponds to the minimum duration required by the control to compensate for the free dynamics of the system.

In the following, after introducing the experimental setup and the main tools of our QOC protocol, we present experimental results, including full state reconstruction after a variable number of Floquet periods, that demonstrate efficient state stabilization of our study case with both methods. The control technique is then extended to states for which the absence of symmetries requires a direct optimization of the Floquet operator.

\section{Experimental setup and numerical algorithms}

\subsection{Experimental setup}
Our experiments are performed on pure $^{87}$Rb BECs of typically $5\cdot10^5$ atoms obtained in an hybrid trap formed by a crossed optical dipole trap and a magnetic quadrupole trap~\cite{fortun_2016}. The BEC is adiabatically loaded into a one-dimensional optical lattice created by two counter-propagating laser beams along the $x$-axis, with a wavelength of $\lambda = 1064$ nm. Along the optical lattice axis, the atoms experience the potential

\begin{equation}
	V(x) = -\dfrac{s}{2}E_\dL \cos\left( k_\dL x + \varphi(t) \right) + V_\text{hyb}(x),
	\label{eq:global_potential}
\end{equation} where $k_\dL=2\pi/d = 2\pi/(\lambda/2)$ and $E_\dL = \hbar^2 k_\dL^2/ 2m$ (with $\hbar$ the reduced Planck constant and $m$ the atomic mass of $^{87}$Rb) are the wavenumber and the characteristic energy associated with the lattice, respectively, with $d$ the lattice spacing. The dimensionless depth of the lattice $s$ is calibrated for each experiment~\cite{cabrera_2018}. 
The hybrid trap potential $V_\text{hyb}(x)$ is characterized by an angular frequency of $\omega_x=2\pi \times 10$ Hz along the lattice axis. In what follows, we neglect this contribution considering the short duration of a typical control ramp. Likewise, the dilute nature of the BEC makes the impact of repulsive interactions on the lattice dynamics negligible. As shown below, the dynamics are therefore governed with a good approximation by a Schr\"odinger equation driven by the lattice potential.

The lattice phase $\varphi(t)$ sets the spatial position of the lattice over time, and is our control parameter. It is varied by setting the relative phase between the drives of two acousto-optic modulators controlling the lattice beams. 
Once loaded into the ground state of the optical lattice, the BEC wavefunction subsequently evolves in the null quasi-momentum subspace and can be written as a superposition of plane waves:
\begin{equation}
	\ket{\Psi}=\sum_{\ell \in \mathbb{Z}} c_\ell \ket{\chi_\ell},
	\label{eq:psi}
\end{equation}
with $c_\ell$ complex coefficients verifying $\sum_{\ell} {\abs{c_\ell}^2}=1$, and $\ket{\chi_\ell}$ the momentum eigenstate with eigenvalue $\ell \hbar k_L$ ($\chi_\ell(x)=e^{i\ell k_L x}/\sqrt{d}$). 
At the end of an experiment, all traps are turned off and the atoms fall for a time-of-flight of $35$ ms, allowing us to access the final in-trap momentum distribution. An absorption image of the atomic distribution after time-of-flight yields equally-spaced diffraction orders from which we extract the populations $\abs{c_\ell}^2$. 

\subsection{Quantum optimal control (QOC) algorithm}
\label{QOC}
The design of the control field follows the strategy presented in~\cite{dupont_2021}. A piecewise constant phase $\varphi(t)=\left\lbrace \varphi(0), \varphi(\Delta t), ..., \varphi{((N-1)\Delta t)} \right\rbrace$ is optimized to prepare a desired target state $\ket{\Psi_\dt}$ from an initial state $\ket{\Psi_0}$ within an optimal control time $t_{\doc}=N\Delta t$ in a lattice with constant depth $s$. Typically $\Delta t\simeq 250\,\mathrm{ns}$ and we use $400$ phase values. To determine the piecewise phase values, we apply a gradient based optimal control algorithm either of first or second order~\cite{fouquieres2011} to iteratively maximize the figure of merit $\mathcal{F}$, which is given by the usual quantum fidelity $\mathcal{F}=|\braket{\Psi(t_{\doc})}{\Psi_\dt}|^2$, where $\ket{\Psi(t_{\doc})}$ is the state obtained from the numerical evolution of $\ket{\Psi_0}$ in a lattice with the phase modulation $\varphi(t)$.

The optimization is performed considering the lattice potential only, and the dynamics is governed by the Hamiltonian (in dimensionless units $\tilde{x}=k_\dL x$, $\tilde{p}=p/\hbar k_\dL$):
 \begin{equation}
 	\tilde{H}=H/E_\dL = \tilde{p}^2 - \frac{s}{2} \cos(\tilde{x}+\varphi(t)).
 	\label{eq:H}
 \end{equation}
From this general procedure, we derive \emph{preparation ramps}, where the initial state $\ket{\Psi_0}$ is the ground state of the lattice, as well as \emph{stabilization ramps} for which $\ket{\Psi_0}=\ket{\Psi_\dt}$. The control duration is chosen to be of order of the period  $T_0$, which is the inverse of the transition frequency between the two lowest lattice energy bands at null momentum for the considered depth $s$ ($T_0\simeq 60\,\upmu$s at $s=5$). This choice results from a compromise: for shorter times, the atomic state does not have time to evolve under the lattice dynamics, and the algorithm cannot converge toward  sufficient fidelity (we generally consider the optimization converged when the fidelity crosses the threshold $\mathcal{F}>0.995$); for very long control times, the algorithm reaches numerical fidelities very close to 1, but experimental fluctuations or imperfections can have a detrimental impact on the actual outcome.

\subsection{State reconstruction}  \label{Reconstruction}
Once a state has been prepared using a QOC ramp, a maximum likelihood reconstruction algorithm can be used to characterize the prepared state. In particular, we can thus verify its experimental fidelity to the target state and its purity. The state reconstruction is performed following the procedure presented in~\cite{dupont_2023}.

The algorithm uses supplemental measurements of the time evolution of the prepared state, held after preparation in a static lattice of fixed depth, and determines from these measurements the density matrix $\hat{\rho}_{\dML}$ maximizing the likelihood function $\mathcal{L}$, $\hat{\rho}_{\dML} = \text{arg\,max}\lbrace{\mathcal{L}\left[\hat{\rho}\right]}\rbrace$. The likelihood is defined as
\begin{equation}
	\mathcal{L}\left[ \hat{\rho} \right] = \prod_{\ell,t} \pi_{\ell,t}^{f_{\ell,t}}, \notag
\label{eq:likelihood}
\end{equation}
where the experimental measurements $f_{\ell,t}=\frac{1}{N_t}\abs{c_\ell(t)}^2$ measure the fraction of atoms in the momentum order $\ell$ at time $t$, with $N_t$ the number of equally spaced time steps. 
The theoretical measurement probabilities $\pi_{\ell,t}=\mathrm{Tr}\left\lbrace \hat{\rho}\hat{E}_{\ell,t} \right\rbrace $  are associated with the operators $\hat{E}_{\ell,t}=\frac{1}{N_t}\hat{U}^\dagger(t,t_{\doc})\ket{\chi_\ell}\bra{\chi_\ell}\hat{U}(t,t_{\doc})$, where $\hat{U}(t,t_{\doc}) $ is the evolution operator in the static lattice between times $t_{\doc}$ and $t$, which altogether form a positive operator-valued measure (POVM).

The likelihood maximization algorithm~\cite{rehacek_2001,lvovsky_2004} iteratively transforms a density matrix, initially chosen proportional to the identity, until it converges toward a fixed point, which is $\hat{\rho}_{\dML}$.
For the results presented in this article, we apply the optimal control and then sample, after preparation, the evolution of the momentum distribution in the static lattice with $\varphi=0$ and same depth $s$, considering the plane-wave components with $|\ell|\leq 4$, over a total time of $100\,\upmu$s with a time step of $5\,\upmu$s ($N_t=21$). Once $\hat{\rho}_{\dML}$ is determined, the fidelity of the experimental preparation to the numerically expected state $F_{\mathrm{exp}}=\bra{\Psi({t_{\doc}})}\hat{\rho}_{\dML}\ket{\Psi({t_{\doc}})}$ and the purity of the reconstructed state $\gamma=\mathrm{Tr}\left( \hat{\rho}_{\dML}^2\right) $ can be calculated.

\section{Stabilization using space and time symmetries}

 \begin{figure}[tbp]
	\includegraphics[width=0.8\linewidth]{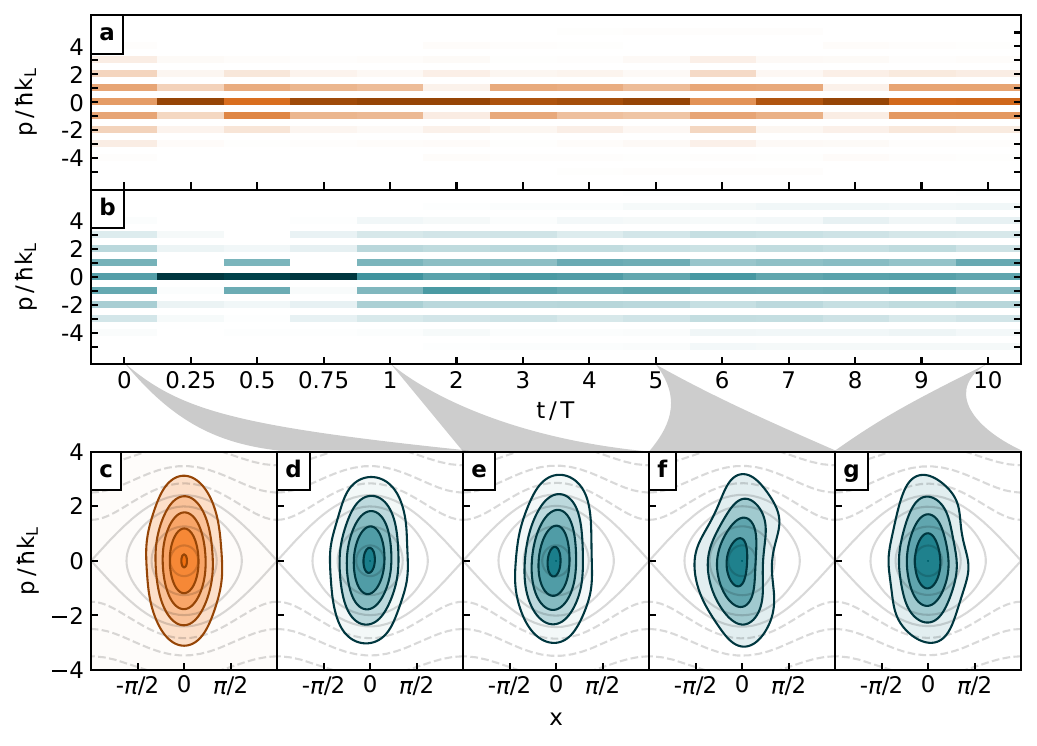}
	\caption{\textbf{\textit{Flip-Flop} stroboscopic stabilization}.  The target state is a Gaussian state centered at the origin of the phase-space of each lattice cell with a squeezing of $\xi=0.5$. The control time is $t_{\doc}=0.86 \, T_0$. (\textbf{a}) Numerical evolution of the momentum distribution. The initial state is prepared from optimal control with a ramp yielding a numerical fidelity $\mathcal{F}=0.996$ in a lattice of dimensionless depth $s = 6.01$.  The first five momentum distributions show the evolution over one period (corresponding to $2 t_{\doc}$ as the stabilization ramp consists in one reverse preparation then one preparation each of duration $t_{\doc}$). The following momentum distributions, measured for an integer number of stabilization periods demonstrate efficient stabilization up to ten periods. (\textbf{b}) Experimentally measured evolution of the momentum distribution in the same conditions as (\textbf{a}), with $s = 6.01 \pm 0.07$. (\textbf{c}) Husimi representation of the initial state, as obtained numerically using the optimal preparation ramp, with fidelity of $\mathcal{F}=0.997$ at $s=5.8$. (\textbf{d}-\textbf{g}) Husimi representations of the experimentally reconstructed stabilized state at different times.  (\textbf{d}) is the reconstructed initial state,(\textbf{e-g}) are the reconstructed Husimi distributions for  1, 5 and 10 periods respectively. The corresponding figures of merit (fidelity and purity) are given in Table~\ref{table:flipflop}.}
	\label{fig:flipflop}
\end{figure}

We choose first, as a case study for $\ket{\Psi_\dt}$ a squeezed Gaussian state $\ket{g\left( x_0, p_0; \xi\right)}$ centered in the phase-space of one lattice cell. The squeezing parameter $\xi$ quantifies the reduction of the position standard deviation $\Delta x = \xi \Delta x_0$ where $\Delta x_0 = k_\dL^{-1} s^{-1/4}$ is the non-squeezed standard deviation at a given depth $s$ (approximately the width of the lattice ground state).
In the null quasi-momentum subspace, this Gaussian state can be expressed on the plane-wave basis as:
\begin{equation*}
	\ket{g( x_0, p_0;\xi)}=\sum_{\ell \in \mathbb{Z}}{c_{\ell}(x_0,p_0;\xi)  \ket{\chi_{\ell}}},
\end{equation*}
where the coefficients are given by
\begin{equation}
	c_\ell(x_0,p_0;\xi) = \left(\dfrac{2\xi^2}{\pi \sqrt{s}}\right)^{1/4} e^{i x_0 p_0 /2} e^{-i \ell x_0} e^{-\xi^2(\ell - p_0)^2/\sqrt{s}}.
	\label{eq:gaussian_states}
\end{equation}
We henceforth consider a state with parameters $(x_0,p_0;\xi)=(0,0;0.5)$ for our case study.

We can first determine with our QOC algorithm a control $\varphi\left( t\right) $ that prepares a state $\ket{\Psi(t_\doc)}$, with a numerical fidelity $\mathcal{F}>0.995$ to the Gaussian target state $\ket{\Psi_\dt}$, starting from the ground state $\ket{\Psi_0}$ of the lattice with a depth $s\simeq 6$, and in a control time $t_{\doc}$.  
Note that this preparation would be difficult to achieve by other means.
As a Gaussian state at the bottom of the lattice well, this squeezed state is analogous to the ground state of a lattice with an effective depth $s_{\mathrm{eff}}=s/\xi^4\sim 100$. This state is technically inaccessible through adiabatic ground state preparation, as it would require too high laser intensities. Moreover, once prepared in the lattice of depth $s$, this state is highly non-stationary and the question of its stroboscopic stabilization then naturally arises.

In a first approach, one can try to reuse the preparation ramp from the ground state, exploiting the state symmetries. Let $\hat{U}$ be the evolution operator generated by the piecewise constant phase $\varphi(t)$ (such that in the ideal case $\hat{U}\ket{\Psi_0}=\ket{\Psi_\dt}$) and $\hat{\Pi}$ and $\hat{\Theta}$ denote the parity and time-reversal operators, respectively.
Then from the properties of the Hamiltonian (\ref{eq:H}), it can be shown that the propagator given by the opposite control, $-\varphi(t)=\left\lbrace -\varphi(0), -\varphi(\Delta t), ..., -\varphi{((N-1)\Delta t)} \right\rbrace$ is $\hat{\Pi} \hat{U} \hat{\Pi}$, and the one given by the time-reversed modulation, $\varphi_{\dflip}(t)=\left\lbrace \varphi((N-1)\Delta t), ..., \varphi(\Delta t), \varphi{(0)} \right\rbrace$ is $\hat{\Theta} \hat{U}^\dagger \hat{\Theta}$. These transformed controls can then be combined to stabilize $\ket{\Psi_\dt}$, depending on its symmetries, with two main cases (see Appendix~\ref{app:transfo} for more details):
\begin{itemize}
\item if both $\ket{\Psi_0}$ and $\ket{\Psi_\dt}$ exhibit time-reversal symmetry of the wavefunction, $\hat{\Theta} \ket{\Psi}=\ket{\Psi}$, one finds:
\begin{equation}
	\ket{\Psi_\dt}=\hat{\Theta}\ket{\Psi_\dt}\xrightarrow{\hat{\Theta}\hat{U}^\dagger\hat{\Theta}}\hat{\Theta}\ket{\Psi_0} = \ket{\Psi_0}\xrightarrow{\hat{U}} \ket{\Psi_\dt}
\end{equation}
which means that a concatenated piecewise control $\varphi_\dT=\left[ \varphi_{\dflip}\left( t\right) , \varphi\left( t\right) \right]$ stabilizes the state $\ket{\Psi_\dt}$. States with this symmetry can be described by a real wavefunction $\Psi^*(x)=\Psi(x)$, and their Husimi representation (see Eq.~\eqref{eq:Husimi} below) is symmetric with respect to the $x$-axis in phase space.
\item if both $\ket{\Psi_0}$ and $\ket{\Psi_\dt}$ exhibit a combined time-reversal and parity symmetry of the wavefunction, $\hat{\Theta}\hat{\Pi} \ket{\Psi}=\ket{\Psi}$ (or equivalently $\hat{\Pi} \ket{\Psi}=\hat{\Theta}\ket{\Psi}$), one finds:
\begin{equation}
	\ket{\Psi_\dt}=\hat{\Theta}\hat{\Pi}\ket{\Psi_\dt}\xrightarrow{\hat{\Theta}\hat{\Pi}\hat{U}^\dagger\hat{\Pi}\hat{\Theta}}\hat{\Theta}\hat{\Pi}\ket{\Psi_0} = \ket{\Psi_0}\xrightarrow{\hat{U}} \ket{\Psi_\dt}
\end{equation}
which means that a concatenated piecewise control $\varphi_\dT=\left[ -\varphi_{\dflip}\left( t\right) , \varphi\left( t\right) \right]$ stabilizes the state $\ket{\Psi_\dt}$. States with this symmetry can be described by a wavefunction that verifies $\Psi^*(x)=\Psi(-x)$, and their Husimi representation is symmetrical about the $p$-axis in phase space.
\end{itemize}
For the squeezed Gaussian state that we have chosen, prepared from the ground state, both $\ket{\Psi_{0,\dt}}$ have time-reversal symmetry. Therefore, given the preparation control $\varphi(t)$, the \emph{flip-flop} control $\varphi_\dT=\left[ \varphi_{\dflip}\left( t\right) , \varphi\left( t\right) \right]$ allows us to stabilize the state with a period $T=2t_{\doc}$.

\begin{table}[ht!]
\begin{center}
\begin{tabular}{
>{\centering}p{0.16\linewidth}
>{\centering}p{0.14\linewidth}
>{\centering}p{0.14\linewidth}
>{\centering}p{0.14\linewidth}
>{\centering\arraybackslash}p{0.14\linewidth}
}
\hhline{=====}
Fig.~\ref{fig:flipflop}&\textbf{d}&\textbf{e}&\textbf{f}&\textbf{g}\\ \hline
$F_{\mathrm{exp}}$&0.96&0.97&0.82&0.86\\
$\gamma$&0.93&0.97&0.8&0.83\\
$s$&$5.8\pm 0.06$&$5.91\pm 0.05$&$5.84\pm 0.06$&$5.74\pm 0.05$ \\\hline
\end{tabular}
\caption{Figures of merit obtained from the reconstruction of the squeezed state stabilized with a \emph{flip-flop} phase ramp (see Fig.~\ref{fig:flipflop}). }
\label{table:flipflop}
\end{center}
\end{table}

In Fig.~\ref{fig:flipflop}, we present the experimental results for such a stabilization approach based on a preparation ramp for the squeezed state $\ket{g(0,0;0.5)}$ in a control time $t_{\doc}=0.86T_0$. The target state is initially prepared by optimal control, then a series of stroboscopic stabilization controls are applied, during which measurements are taken to characterize the system evolution. Panels (\textbf{a-b}) show the numerical and experimentally measured time evolution of the plane-wave populations, respectively. The first five images depict the evolution over one period $T=1.72T_0$: as expected, we recover the ground state distribution at the intermediate time $t/T=0.5$. At this time, the reversed control $\varphi\left( -t\right)$ has effectively undone the preparation. At $t/T=1$, the distribution is similar to the first image, representing a momentum distribution of a state close to the target.

The density matrix of the state obtained after a varying number of periods of stroboscopic stabilization is reconstructed (see Sec.~\ref{Reconstruction}) to characterize the state purity and its fidelity to the target during the stabilization process. The resulting Husimi distributions of the maximum-likelihood density matrices
\begin{equation}
Q_{[\hat{\rho}_\mathrm{ML}]}(x,p)\equiv \frac{1}{2\pi}\bra{g(x,p;1)}\hat{\rho}_\mathrm{ML}\ket{g(x,p;1)} \label{eq:Husimi}
\end{equation}
can be computed and represented on the phase space of a lattice cell.
Figure \ref{fig:flipflop}(\textbf{c}) shows the Husimi distribution of the prepared numerical state obtained using $\varphi\left( t\right)$, while (\textbf{d-g}) represent the distributions deduced from experimental reconstructions for $t/T = 0, 1, 5$ and $10$ periods. Although the Husimi distributions exhibit some small variations compared to the prepared state, they remain very similar stroboscopically, demonstrating the efficient stabilization of the highly non-stationary Gaussian squeezed state. We quantify the efficiency of the stabilization process from the fidelity to the numerically prepared state, and the purity of the reconstructed state (see Table~\ref{table:flipflop}), and these quantities are larger than 0.8 for all the measurement times considered. In order to account for small fluctuations of the experiment, the lattice depth was systematically re-calibrated for each panel presenting experimental data, and the control ramps slightly re-optimized. Specific lattice depth values are therefore also indicated in Table~\ref{table:flipflop}.

\section{A generic optimal stabilization scheme}

The previous use of symmetry allows a stroboscopic stabilization, but only for a fraction of accessible states. In the general case, where the target state does not necessarily have any time-reversal or parity symmetry, we can resort to optimal control for a direct search of a stabilizing control.
Using our QOC protocol (see section Sec.~\ref{QOC}) we search for an optimal phase modulation that transforms the target state into itself in a finite fixed control time $t_{\doc}$. From this control procedure, the state of the system is allowed in principle to explore the full Hilbert space, and this \textit{target-to-target} approach does not require the system to reach any specific state, with a given symmetry, at half-period. The control is thus expected to be shorter because of the single constraint to satisfy, as opposed to the concatenated control presented above, which effectively imposes two constraints in a stabilization period.

  \begin{figure}[tbp]
	\includegraphics[width=0.8\linewidth]{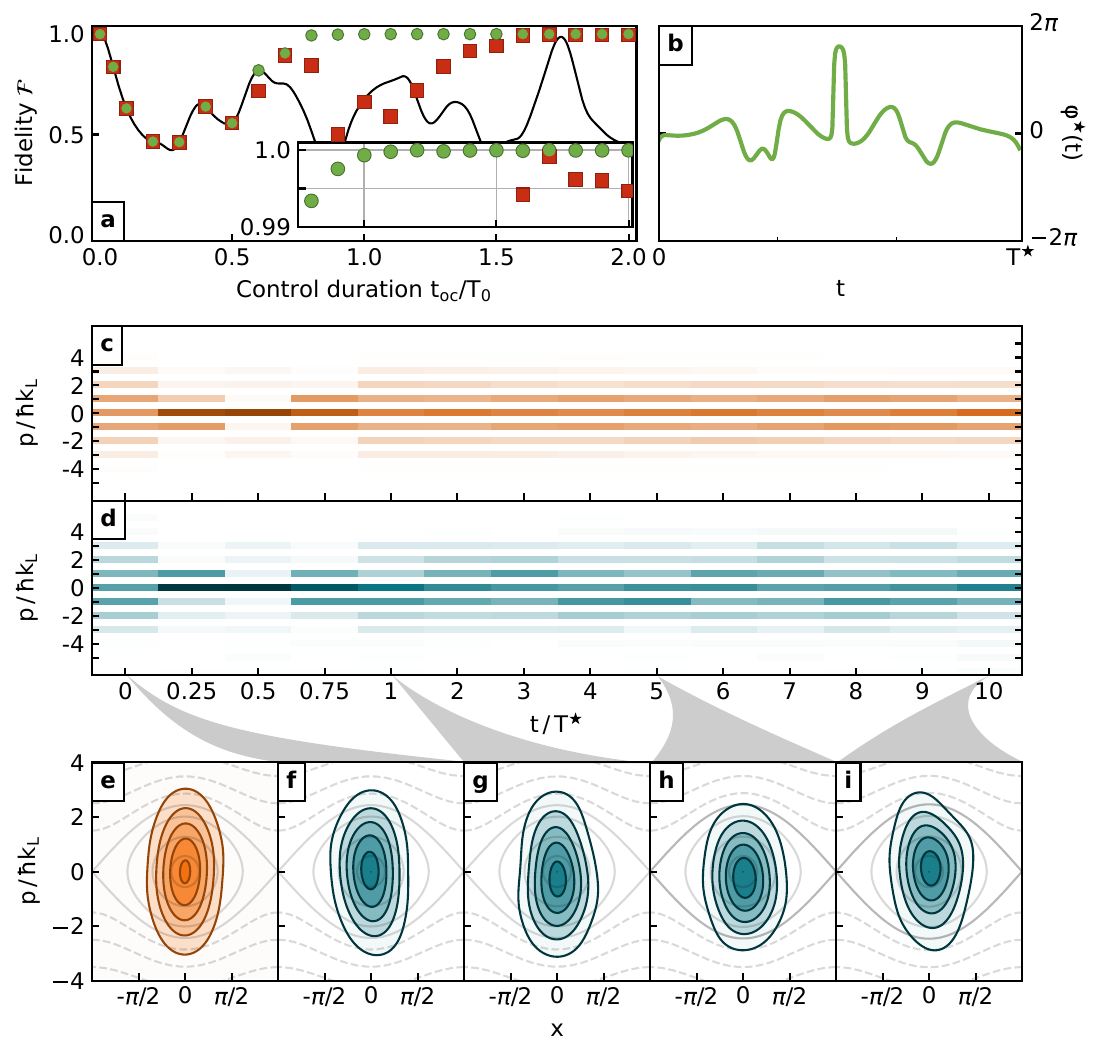}
	\caption{\textbf{Direct \textit{target-to-target} stabilization and minimum time.}  (\textbf{a}) Numerical comparison of the performance of the \textit{flip-flop} and \textit{target-to-target} methods with the evolution in a static lattice: optimal numerical fidelity for stabilization over one period, depending on the control time $t_{\doc}$, starting from $\ket{\Psi_\dt}=\ket{g(x_0=0,p_0=0;\xi=0.5)}$ for $s=5.5$. The black line is the evolution of the state in the static lattice ($\varphi(t)=0$), the red squares are the result of an optimized preparation and \textit{flip-flop} stabilization and the green circles are the result of optimal \textit{target-to-target} stabilization.  (\textbf{b})  The optimal phase modulation $\varphi^\star\left( t\right) $ obtained to prepare the squeezed Gaussian state from itself in a depth $s=5.92$ with fidelity $\mathcal{F}>0.995$ during the minimum time control $T^\star=0.86T_0$. (\textbf{c}) (resp. (\textbf{d})) Numerical (resp. experimental) evolution of the momentum distribution after preparation of the initial state, using the phase ramp plotted in (\textbf{b}) to hold the state stroboscopically during ten periods in a lattice of depth $s=6.04\pm 0.06$. 
The first five momentum distributions show the evolution over one period, and the following ones demonstrate efficient stabilization up to ten periods. (\textbf{e}) Numerical Husimi representation of the initially prepared state obtained using an optimal ramp, with fidelity $\mathcal{F}>0.995$, in a depth $s=5.94$. (\textbf{f}-\textbf{i}) Husimi representations of the reconstructed stabilized state at different times.  (\textbf{f}) is the experimental reconstructed initial state,(\textbf{g-i}) are the reconstructed Husimi distributions for  1, 5 and 10 periods respectively. The corresponding figures of merit are given Table~\ref{table:psipsi}.		
	 } 
	\label{fig:compare}
\end{figure}

In Fig.~\ref{fig:compare}(\textbf{a}), we compare the stroboscopic fidelity after one period obtained from the optimal (numerical) phase control, as a function of total control time, for both \emph{flip-flop} and \emph{target-to-target} methods, to stabilize the out-of-equilibrium state $\ket{\Psi_\dt}=\ket{g(x_0=0,p_0=0;\xi=0.5)}$ in a lattice of depth $s=5.5$. This result is also compared to the fidelity obtained from a free evolution in the static lattice potential.
One can first notice that for the state and depth studied here, a "revival" of the state occurs in the free evolution in the lattice, at around $1.75\,T_0$. This is an accidental phase synchronization of the lattice eigenstates decomposing the target state. This phenomenon is neither generic, nor periodic (as opposed \emph{e.g.} to what would  happen in an harmonic oscillator), and there is no certainty on the time at which it may occur or the value of the stroboscopic fidelity that may be reached this way for other depth and squeezing parameters. 

Both approaches of optimal control achieve a better fidelity more reliably.
One observes that, for very short control durations, both methods fail to achieve a better result than the free evolution in a static lattice: this highlights that there is a minimum time related to the natural dynamics of the system below which optimal control cannot be achieved. In other words, for such short times the phase modulation cannot compensate the field-free evolution of the system. This originates from the inertia of the atoms in the lattice, and happens below a timescale of order $T_0$ (which scales as $T_0\sim 1/(\sqrt{2s}\nu_\dL)$ for $s$ large, with $\nu_\dL=E_\dL/h$ the lattice characteristic frequency).
As expected, we then see that the \textit{target-to-target} method yields fidelities larger than $0.995$ for shorter control times ($<T_0$) compared to the \textit{flip-flop} method. A time control of $T^\star=0.86\,T_0$ is enough for this threshold, and almost the double duration is required with the \textit{flip-flop} method. In an intermediate regime, the \emph{flip-flop} method may even perform worse than the free evolution, as only the preparation ramp is optimized.

Figure~\ref{fig:compare}(\textbf{b}) shows the stabilization ramp obtained for a time $T^\star$ and (\textbf{c}) (resp. (\textbf{d})) the corresponding numerical (resp. experimental) evolution after preparation of the state with a distinct control ramp at $t=0$ and the following periodic application of the stabilization control. 
As in the previous case, the state can be reconstructed during the evolution, after a variable number of periods $t/T^\star=0,1,5,10$, and the results are presented in Fig.~\ref{fig:compare}(\textbf{e-i}). The figures of merit from these reconstructions are listed in Table~\ref{table:psipsi}: fidelities of the prepared state are always greater than 0.9, and the purity of the reconstruction higher than 0.88.

In obtaining a control ramp that stabilizes the squeezed Gaussian state, we effectively engineer a Floquet unitary operator for which this state is (with a good approximation) an eigenstate. Residual variations on the experimental data can be interpreted as originating from interferences with small amplitudes on Floquet states with different quasi-energies
\footnote{If the maximum overlap of the target state with a Floquet state $\ket{\phi_0}$ is smaller than 1, $|\braket{\Psi_\dt}{\phi_0}|^2=1-\eta$, $\eta\ll1$ , the target state can be written $\ket{\Psi_\dt}=\sqrt{1-\eta}\ket{\phi_0}+\sqrt{\eta}\sum_{k>1} c_k\ket{\phi_k}$, with $\{\ket{\phi_k}\}$ the Floquet states of the evolution operator with quasi-energies $\epsilon_k$, and the state after an evolution of $n$ periods $T$ can be written as:
\[\ket{\Psi(nT)}\propto\sqrt{1-\eta}\ket{\phi_0}+\sqrt{\eta}\sum_{k>1} c_ke^{-i(\epsilon_k-\epsilon_0) nT}\ket{\phi_k}\]}.

The optimal control search highlights the existence of a minimum time $T^\star$ for which the stabilization can be achieved. A quantum speed limit can thus be defined for this complex control problem~\cite{Deffner2017,Caneva2009}, here more precisely for the exact transfer of a delocalized state of the lattice to itself. This minimal time is close to the natural timescale of the dynamics, as the control inherently depends on the inertial behavior of the atoms in the lattice, a scaling also found in other quantum brachistochrone problems
\footnote{This minimum time can be compared to an estimation of the lower bound that is independent of the specific shape of the control parameter $\varphi(t)$, such as, e.g., the Mandelstam and Tamm derivation~\cite{Deffner2017}. Using in first approximation the fact that the control has time reversal symmetry, this bound can be roughly estimated as $$
T_{\textrm{QSL}}=\frac{2\hbar\arccos(|\langle\Psi_\dt|\Psi(T^\star/2)\rangle|)}{\overline{\Delta H}}
$$
where the variance $\overline{\Delta H}=\overline{\langle H^2\rangle-\langle H\rangle ^2}$ is a time-averaged value over the control duration.
In the case of Fig.~\ref{fig:compare}, we obtain $T_{\textrm{QSL}}\simeq 0.15 T_0$. This is indeed smaller than $T^\star=0.86 T_0$, but it highlights the impact that the detailed dynamics can have on the limits of the control.}~\cite{Lam2021}.

\begin{table}[ht!]
\begin{center}
\begin{tabular}{
>{\centering}p{0.16\linewidth}
>{\centering}p{0.14\linewidth}
>{\centering}p{0.14\linewidth}
>{\centering}p{0.14\linewidth}
>{\centering\arraybackslash}p{0.14\linewidth}
}
\hhline{=====}
Fig.~\ref{fig:compare}&\textbf{f}&\textbf{g}&\textbf{h}&\textbf{i}\\ \hline
$F_{\mathrm{exp}}$&0.98&0.97&0.92&0.91\\
$\gamma$&0.98&0.97&0.89&0.88\\
$s$&$5.94\pm 0.06$&$5.93\pm 0.09$&$6.03\pm 0.07$&$6.08\pm 0.07 $ \\\hline
\end{tabular}
\caption{Figures of merit obtained from the reconstruction of the squeezed state stabilized with a \emph{target-to-target} phase ramp (see Fig.~\ref{fig:compare}). }
\label{table:psipsi}
\end{center}
\end{table}

\begin{figure}[tbp]
	\includegraphics[width=0.8\linewidth]{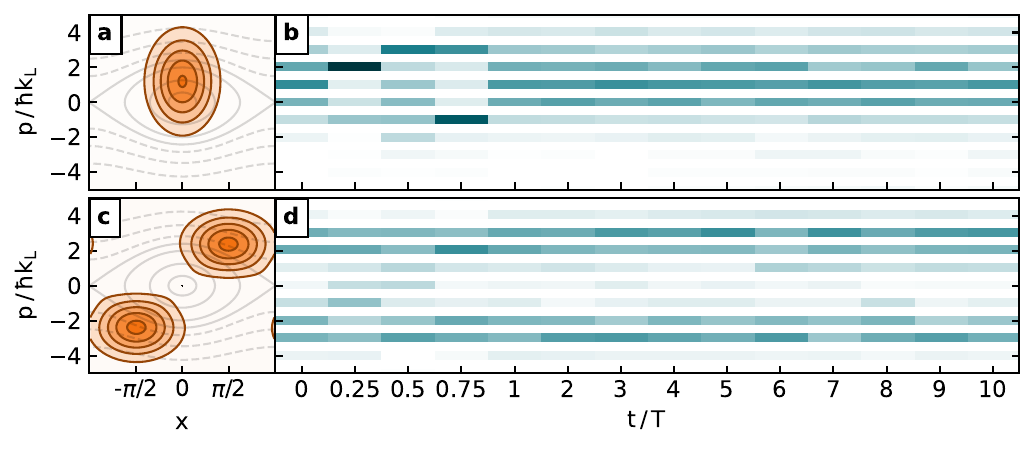}
	\caption{\textbf{Other stroboscopic stabilizations.} (\textbf{a}) Husimi distribution of a Gaussian state with squeezing $\xi=0.5$ centered on $x_0=0, p_0=\sqrt{s}/2 $ in the phase-space for $s=5.84$. (\textbf{b})  Corresponding experimental stabilization during ten periods, in a lattice of depth $s=5.84 \pm 0.08$. (\textbf{c}) Husimi distribution of a superposition of two non-squeezed Gaussian states centered in $x_0=\pm \pi/2, p_0=\pm \sqrt{s} $ in the phase-space for $s=5.8$. (\textbf{d}) Corresponding experimental stabilization during ten periods, in a lattice of depth $s=5.8 \pm 0.07$.}
	\label{fig:ExoticStates}
\end{figure}

Finally, the direct \emph{target-to-target} control method allows for the stabilization of states that do not exhibit specific symmetries. This is demonstrated by the results shown in Fig.~\ref{fig:ExoticStates}: a first considered target state is a Gaussian state with squeezing $\xi=0.5$, centered at $x_0=0$ and $p_0=\sqrt{s}/2$. The associated Husimi distribution is represented in panel (\textbf{a}), and the experimental results over ten periods for $T=1.5 T_0$ are presented in (\textbf{b}). The measurements show that the distribution remains nearly identical over the observation times. Similarly, a state corresponding to a superposition of two non-squeezed Gaussian states $\ket{g\left( \pm \pi/2, \pm \sqrt{s}; 1\right)}$, for which the preparation ramp cannot be re-used, is represented in (\textbf{c}). The experimental results for stabilizing this state with a period $T=0.86 T_0$ are shown in (\textbf{d}). Once again, the distribution remains quasi-stationary at stroboscopic times, demonstrating the efficiency of our stabilization method.

\section{Conclusion}

In this work we have presented how quantum optimal control can achieve the stroboscopic stabilization of quantum states, in the context of a Bose-Einstein condensate in an optical lattice potential. An optimal control ramp can be derived from a state preparation ramp under certain conditions of symmetry, otherwise a direct optimization is possible. In the latter case, numerical calculations show a smooth transition, as a function of the control time, from a regime where the control cannot be optimized to a regime of efficient state stabilization, highlighting the existence of a minimal time or quantum speed limit. Our experimental results demonstrate the efficiency of the protocol for the stabilization of a variety of states.

In the continuation of these results, an interesting perspective is to further characterize the quantum speed limit and the corresponding solutions: we indeed observe that while several controls perform equivalently when the control time is long, the optimization for the minimum time $T^\star$ mostly converges to a unique control shape, with an apparent time-reversal symmetry (see Fig.~\ref{fig:compare}(\textbf{b})), in spite of the latter not being a constraint. This work could also be extended to the simultaneous stabilization of several Floquet states, or a Floquet subspace with a degenerate quasi-energy. This latter extension has the advantage of stroboscopically stabilizing any state of this subspace. Through this approach, the search of effective Hamiltonian~\cite{goldman_dalibard_2014} can be re-framed as the optimization of the evolution of states: as an example in the case of driven ratchet transport in a modulated lattice~\cite{ratchet_2023}, the transport of the lattice ground state can be directly optimized, rather than depend on the variation of parameters of a harmonic modulation. 

\section{Acknowledgments}

This work was (partially) supported through the EUR Grant NanoX No. ANR-17-EURE-0009 in the framework of the ``Programme d’Investissements d’Avenir'' and research funding Grant No. ANR-22-CE47-0008. F.A. and N.D. acknowledge support from R\'egion Occitanie and Universit\'e Toulouse III-Paul Sabatier. N.O. acknowledges support from from R\'egion Occitanie and Thales Alenia Space.
 
\begin{appendix}
\section{Evolution under time and parity-reversed controls}
\label{app:transfo}

We denote by $\hat{\Pi}$ the parity operator, the linear operator defined by its action in the position or momentum bases:
\[\hat{\Pi}\ket{x}=\ket{-x}, \quad \hat{\Pi}\ket{p}=\ket{-p}.\]
It is its own inverse, and its action on the position and momentum operators is given by:
\[\hat{\Pi}\hat{X}\hat{\Pi}=-\hat{X}, \quad \hat{\Pi}\hat{P}\hat{\Pi}=-\hat{P}.\]
From the definition of the Husimi distribution~\eqref{eq:Husimi}, a density matrix invariant under the action of parity $\hat{\Pi}\hat{\rho}\hat{\Pi}=\hat{\rho}$ has a Husimi distribution with central symmetry.

From the Hamiltonian~\eqref{eq:H}, the action of parity corresponds to a change of $\varphi(t)$ into $-\varphi(t)$. Therefore we can deduce the action of parity on the evolution operator corresponding to the piecewise constant phase $\varphi(t)=\{\varphi(0),\varphi(\Delta t),...\varphi((N-1)\Delta t)\}$:
\begin{align*}
\hat{\Pi}\hat{U}[\varphi(t)]\hat{\Pi}&=\hat{\Pi}\prod_{j=0}^{N-1}\exp\left(-i\hat{\tilde{H}}\left[\varphi(j\Delta t)\right]\Delta t\right)\hat{\Pi}\\
&=\prod_{j=0}^{N-1}\hat{\Pi}\exp\left(-i\hat{\tilde{H}}\left[\varphi(j\Delta t)\right]\Delta t\right)\hat{\Pi}\\
&=\prod_{j=0}^{N-1}\exp\left(-i\hat{\tilde{H}}\left[-\varphi(j\Delta t)\right]\Delta t\right)\\
&=\hat{U}[-\varphi(t)]
\end{align*}

\medskip

Likewise, we denote by $\hat{\Theta}$ the time reversal operator, the anti-unitary operator which can be defined by its action on the wavefunction for the external degree of freedom:
\[\bra{x}\hat{\Theta}\ket{\Psi}=\braket{\Psi}{x}^*,\]
corresponding to complex conjugation.
The position and momentum operators are transformed as:
\[\hat{\Theta}\hat{X}\hat{\Theta}=\hat{X}, \quad \hat{\Theta}\hat{P}\hat{\Theta}=-\hat{P}.\]
From the definition of the Husimi distribution~\eqref{eq:Husimi}, a density matrix invariant under time reversal $\hat{\Theta}\hat{\rho}\hat{\Theta}=\hat{\rho}$ has a Husimi distribution with axial symmetry with respect to the $x$-axis. Furthermore the Husimi distribution of a state exhibiting the combined parity-time reversal symmetry $\hat{\Theta}\hat{\Pi}\hat{\rho}\hat{\Pi}\hat{\Theta}=\hat{\rho}$ has an axial symmetry with respect to the $p$-axis.

The operator $\hat{\Theta}$ is its own inverse, and leaves the control Hamiltonian~\eqref{eq:H} invariant. However due to the antilinearity of $\hat{\Theta}$, its action  on the evolution operator corresponding to the piecewise constant phase $\varphi(t)=\{\varphi(0),\varphi(\Delta t),...\varphi((N-1)\Delta t)\}$ is:
\begin{align*}
\hat{\Theta}\hat{U}[\varphi(t)]\hat{\Theta}&=\hat{\Theta}\prod_{j=0}^{N-1}\exp\left(-i\hat{\tilde{H}}\left[\varphi(j\Delta t)\right]\Delta t\right)\hat{\Theta}\\
&=\prod_{j=0}^{N-1}\hat{\Theta}\exp\left(-i\hat{\tilde{H}}\left[\varphi(j\Delta t)\right]\Delta t\right)\hat{\Theta}\\
&=\prod_{j=0}^{N-1}\exp\left(i\hat{\tilde{H}}\left[\varphi(j\Delta t)\right]\Delta t\right)\\
&=\hat{U}^\dagger[\varphi_{\dflip}(t)]
\end{align*}

We can conclude from the preceding results that:
\begin{itemize}
\item the application of the time-reversed phase control $\varphi_{\dflip}(t)$ yields an evolution operator $\hat{U}[\varphi_{\dflip}(t)]=\hat{\Theta}\hat{U}^\dagger[\varphi(t)]\hat{\Theta}$,
\item the application of the time-reversed phase control with the opposite sign $-\varphi_{\dflip}(t)$ yields an evolution operator $\hat{U}[-\varphi_{\dflip}(t)]=\hat{\Theta}\hat{\Pi}\hat{U}^\dagger[\varphi(t)]\hat{\Pi}\hat{\Theta}$,
\end{itemize}
both results being used in the main text.

\end{appendix}

\bibliography{article_medaille_bib}

\end{document}